\documentclass[runningheads,fleqn]{svmult}
\usepackage{makeidx}   
\usepackage{graphicx}  
\usepackage{subeqnar}  
\usepackage{multicol}  
\usepackage{taphys}    
\makeindex             
\usepackage{amsmath}   
\mathindent\parindent 
\newcommand{\imai}{{\rm i}}
\newcommand{\e}{{\rm e}}
\renewcommand{\d}{{\rm d}}
\usepackage{bm}
\begin{document}

\title*{
{\rm \footnotesize  to appear in {\em Advances in Solid State
 Physics}, ed. by B. Kramer (Springer 2003)}\\
Simulation of Transport and Gain in Quantum Cascade Lasers}
\toctitle{Simulation of transport and gain in quantum cascade lasers}
\titlerunning{Transport and gain in quantum cascade lasers}
\author{A. Wacker\inst{1}
\and S.-C. Lee\inst{1}
\and M.F. Pereira Jr.\inst{1,2}}
\authorrunning{A. Wacker, S.-C. Lee, and M.F. Pereira Jr.}
\institute{Institut f{\"u}r Theoretische Physik, Technische Universit{\"a}t
Berlin, Hardenbergstr. 36, 10623 Berlin, Germany
\and NMRC, Lee Maltings, Prospect Row, Cork, Ireland
}
\maketitle

\begin{abstract}
Quantum cascade lasers can be modeled within a hierarchy of different
approaches: Standard rate equations for the electron densities in
the levels, semiclassical Boltzmann equation for the microscopic
distribution functions, and quantum kinetics including the
coherent evolution between the states.
Here we present a quantum transport approach based on 
nonequilibrium Green functions. This allows for quantitative 
simulations of the transport
and optical gain of the device. The division of the current density
in two terms shows that semiclassical transitions are likely
to dominate the transport for the prototype device of Sirtori {\it et al.}
but not for a recent THz-laser with only a few layers per period.
The many particle effects are extremely dependent on the design of the 
heterostructure, and for the case considered here, 
inclusion of electron-electron interaction at the Hartree Fock level, 
provides a sizable change in absorption but imparts only
a minor shift of the gain peak.
\end{abstract}

\section{Introduction}
Since the first realization of a quantum cascade laser (QCL) in 1994
\cite{FAI94a} these semiconductor heterostructures have become 
important devices in the infrared regime operating up to
room temperature \cite{BEC02a}.  Lasing in the THz-region was 
also achieved recently \cite{KOE02}, opening a new
window for applications. The standard devices contain an
injector region guiding the electrons to the upper laser level
in an active region where the optical transitions occur between
a few discrete levels. A frequently studied prototype is the sample
in \cite{SIR98}.
Different designs are interminiband-QCLs \cite{SCA97,STR99}, as well as QCLs 
without injector regions \cite{WAN01} or containing only four
barriers per period like the staircase-laser \cite{ULB02} and a
recent THz-QCL \cite{WIL03}.

The modeling of quantum cascade lasers was first performed
on the basis of rate equations \cite{CAP96} for the electron
dynamics in the active region. It was assumed that the
electrons reach the upper laser level with the rate
$J/e$, where $J$ is the current density and $e<0$ is the  
electron charge. A necessary condition for inversion is
that the scattering rate $1/\tau_{u\to l}$ from the upper
to the lower laser level is  smaller than the out scattering
$1/\tau_l$ from the lower laser level. Optimizing these 
scattering rates by a sophisticated choice of well
and barrier widths in the active region, QCLs with high 
performance could be designed. While typically scattering
with optical phonons is considered to be the main mechanism
for the scattering rates \cite{PAU98,SLI99},
electron-electron scattering has also been treated \cite{HYL96,HAR99}. 
The influence of a magnetic field has been studied by these rate equations 
in \cite{APA01}.
While these rate equations for the
electron densities $n_i$ [in units 1/cm$^2$] for the levels $i$
average over the momentum ${\bf k}$ in the in-plane direction,
the distribution functions $f_i({\bf k})$ can be 
taken into account employing Monte-Carlo (MC)
simulations \cite{TOR99,IOT00,IOT01}.

If one includes the injector region in the simulation and imposes
periodic boundary conditions (a good approximation as typical
devices have approximately 30 periods each containing an active region
and an injector region) a full simulation of
QCL-devices can be performed. Such an approach was performed
almost simultaneously on the basis of rate equations  \cite{DON01},
MC-simulations \cite{IOT01a} and a quantum transport
model \cite{WAC01a} obtaining good results for the current-voltage
characteristic of a prototype device \cite{SIR98}.

In this article we want to show, in how far quantum effects affect
the transport and gain behavior and address the question if 
simple semiclassical models such as rate equations or MC-simulations 
are applicable.
In particular we demonstrate (i) that the current can be
calculated in a quantum transport model, (ii) how this relates
to semiclassical approaches, and (iii) discuss the 
implications of many particle corrections on the gain spectra.

\section{Current in Quantum Transport}

In order to describe the quantum cascade laser we start
by defining a set of single particle basis states
$\Psi_{\alpha}(z)\e^{\imai {\bf k}\cdot {\bf r}}/\sqrt{A}$.
Here ${\bf k},{\bf r}$ are two dimensional vectors in
the $x,y$ plane perpendicular to the growth direction $z$
and $A$ is the normalization area.
The functions $\Psi_{\alpha}(z)$ reflect the 
layer sequence of the QCL structure and may be chosen
as energy-eigenstates or Wannier states (see the discussion
in  \cite{LEE02a}). Then the 
Hamilton operator reads in second quantization:
\begin{equation}
\hat{H}=\underbrace{\sum_{\alpha,\beta,{\bf k},s}
H^o_{\alpha\beta}({\bf k}) a_{\alpha,{\bf k}}^{\dag}a_{\beta,{\bf k}}}_{
\hat{H}^o}+\hat{H}_{\rm scatt}
\end{equation}
where all terms connecting different ${\bf k}$-indices (i.e. breaking
the translational invariance of the structures) have been included in
$\hat{H}_{\rm scatt}$. The spin index $s$ yields an additional factor $2$
for the current and the gain 
as we assume that all states are spin degenerate and no spin transitions
occur.

Much information is contained in the
(reduced) density matrix
\begin{equation}
\rho_{\alpha{\bf k},\beta{\bf k}'}=
\langle \hat{a}_{\beta,{\bf k}'}^{\dag}\hat{a}_{\alpha,{\bf k}}\rangle\, ;
\end{equation}
in particular the occupation probabilities
are given by the  diagonal elements $f_{\alpha}({\bf k})=
\rho_{\alpha{\bf k}\alpha{\bf k}}$.
The average current density (in the $z$-direction) is evaluated by
the temporal evolution of the position  operator $\hat{z}$ 
\begin{equation}
J=\frac{e}{V}\left\langle \frac{\d}{\d t}\hat{z}\right\rangle
=\underbrace{\frac{e}{V}
\frac{\imai}{\hbar}\langle[\hat{H}^o,\hat{z}]\rangle}_{=J_0}
+\underbrace{\frac{e}{V}\frac{\imai}{\hbar}
\langle[\hat{H}_{\textrm{scatt}},\hat{z}]
\rangle}_{=J_{\textrm{scatt}}}\, ,
\end{equation}
where $V$ denotes the normalization volume.
Let us first consider the current $J_0$. For an arbitrary choice
of the basis we may write
\begin{equation}
J_0=2\textrm{(for Spin)}\frac{e}{V}\sum_{\alpha\beta{\bf k}}
\frac{\imai}{\hbar}W_{\beta,\alpha}({\bf k})
\rho_{\alpha{\bf k},\beta{\bf k}}
\end{equation}
where
\begin{equation}
W_{\beta,\alpha}({\bf k})=\sum_{\gamma} 
H^o_{\beta\gamma}({\bf k})z_{\gamma\alpha}-z_{\beta\gamma}
H^o_{\gamma\alpha}({\bf k})
\label{EqW}
\end{equation}
is an anti-hermitian matrix. If the wave functions $\Psi_{\alpha}(z)$
are chosen real, which is typical for bound states, 
$W_{\beta,\alpha}({\bf k})$ becomes real and  $J_0$ is determined
by the non-diagonal elements of $\rho_{\alpha{\bf k},\beta{\bf k}}$.

For a scattering part of the form
\begin{equation}
\hat{H}_{\rm scatt}=\sum_{{\alpha\gamma{\bf k, k'},s}}
\hat{O}_{\alpha{\bf k},\gamma{\bf k'}}(t)
\hat{a}^\dagger_{\alpha{\bf k}}(t)\hat{a}_{\gamma{\bf k'}}(t)\ ,
\end{equation}
which contains only pairs of electronic
annihilation and creation operators, we obtain
\begin{equation}
J_{\rm scatt} 
 = \frac{2e}{V\hbar} 
\sum_{\alpha{\bf k}}
\sum_{\gamma\beta {\bf k'}} \imai
 \left \langle \hat{a}^\dagger_{\alpha{\bf k}} 
          \left[\hat{O}_{\alpha{\bf k},\beta{\bf k'}}(t) z_{\beta\gamma} 
           \,-\, z_{\alpha\beta}\hat{O}_{\beta{\bf k},\gamma{\bf k'}}
            \right] 
           \hat{a}_{\gamma{\bf k'}} \right\rangle\, .
\label{EqJscatt}
\end{equation}
In the case of phonon scattering $\hat{O}_{\alpha{\bf k},\gamma{\bf k'}}$ 
contains phonon annihilation and creation operators and thus
phonon-assisted density matrices \cite{KUH98} 
determine $J_{\rm scatt}$ for this scattering process.

To evaluate the density matrices we perform the perturbation expansion
within the formalism of nonequilibrium Green functions
\cite{KAD62,KEL65,HAU96} similar to  \cite{WAC02}.
The key quantities are the lesser and retarded Green function
\begin{eqnarray}
G^<_{\alpha_1,\alpha_2}({\bf k};t_1,t_2)&=&
\imai\langle a^{\dag}_{\alpha_2 {\bf k}}(t_2)a_{\alpha_1 {\bf k}}(t_1)\rangle\\
G^{\rm ret}_{\alpha_1,\alpha_2}({\bf k};t_1,t_2)&=&
-\imai\Theta(t_1-t_2)\langle 
a_{\alpha_1 {\bf k}}(t_1)a^{\dag}_{\alpha_2 {\bf k}}(t_2)+ 
a^{\dag}_{\alpha_2 {\bf k}}(t_2)a_{\alpha_1 {\bf k}}(t_1)
\rangle
\end{eqnarray}
where the time dependence is taken in the Heisenberg picture. 
The lesser Green function refers to the electron density and
it becomes the density matrix
$\rho_{\alpha_1{\bf {\bf k}},\alpha_2{\bf {\bf k}}}(t)=
G^<_{\alpha_1,\alpha_2}({\bf k};t,t)/\imai$ for equal times.
In the stationary state considered here
the Green functions only depend on the time difference $t=t_1-t_2$
and we introduce the energy $E$ as the Fourier conjugate of
$t$:
\begin{equation}
G(t_2+t,t_2)=
\int \frac{\d E}{2\pi}G(E)\e^{-\imai Et/\hbar}
\end{equation}
This provides us with the 
Dyson equation
\begin{equation}
\left(E-{\bf H}^o({\bf k})-\pmb{\Sigma}^{\rm ret}({\bf k},E)\right)
{\bf G}^{\rm ret}({\bf k},E)={\bf 1}
\end{equation} 
and the Keldysh relation
\begin{equation}
{\bf G}^{<}({\bf k},E)={\bf G}^{\rm ret}({\bf k},E)\pmb{\Sigma}^{<}({\bf k},E)
{\bf G}^{\rm adv}({\bf k},E)
\end{equation}
where capital bold symbols represent matrices in $\alpha\beta$.
Together with the functionals $\pmb{\Sigma}\{{\bf G}\}$ for 
the self-energies 
this provides a self consistent set of equations which can be solved
numerically. 
Although the Green functions are diagonal in ${\bf k}$, 
the expression (\ref{EqJscatt}) for $J_{\textrm{scatt}}$
can be evaluated by Eq.~(\ref{EqJscattGF}) as derived 
in the appendix.

Here we use self-energies in self-consistent Born approximation
for impurity, interface roughness, 
and phonon scattering, applying the following
approximations:
(i) The ${\bf k}$-dependence of the scattering matrix elements 
is neglected.
(ii) It is assumed  that
$\pmb{\Sigma}$ is diagonal and depends only on the diagonal elements
of ${\bf G}$ in the basis of Wannier functions.
The scattering matrix elements are evaluated for a typical momentum 
transfer assuming an interface roughness with
average height of $0.28/\sqrt{2\pi}$ nm \footnote{In the
calculations performed in \cite{LEE02a} 
a factor $2\pi$ was lacking in the program, which can be compensated
by the reduction in the roughness height.}
and a correlation length of 10 nm. The impurity scattering
was estimated by an effective scattering rate $\gamma_{\rm imp}/\hbar$.
Electron-electron interaction is included within the
mean field approximation. See  \cite{LEE02a} for further 
details.

\section{The Current-Voltage Characteristic}

\begin{figure}[t]
\includegraphics[width=.48\textwidth]{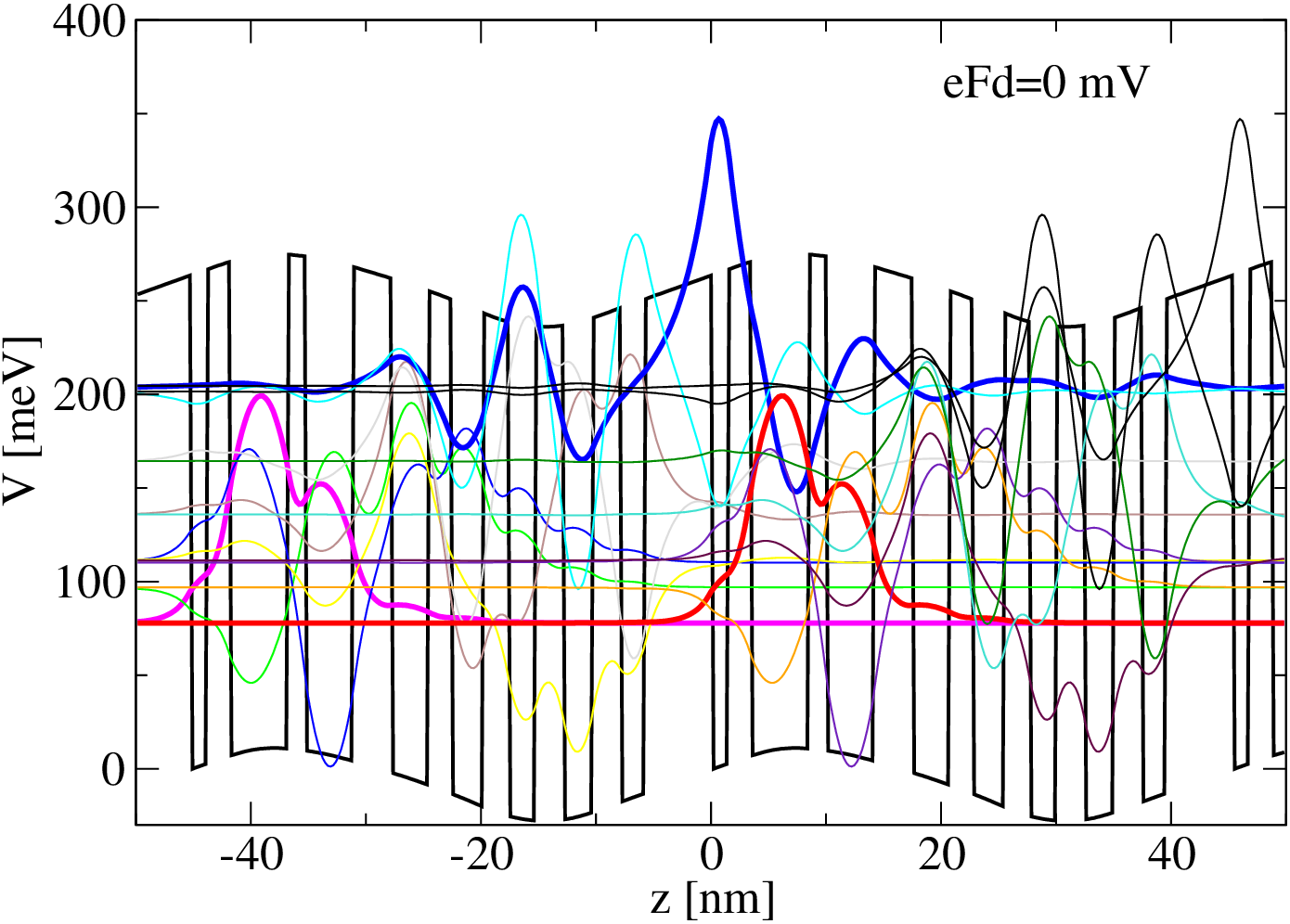}
\hfill 
\includegraphics[width=.48\textwidth]{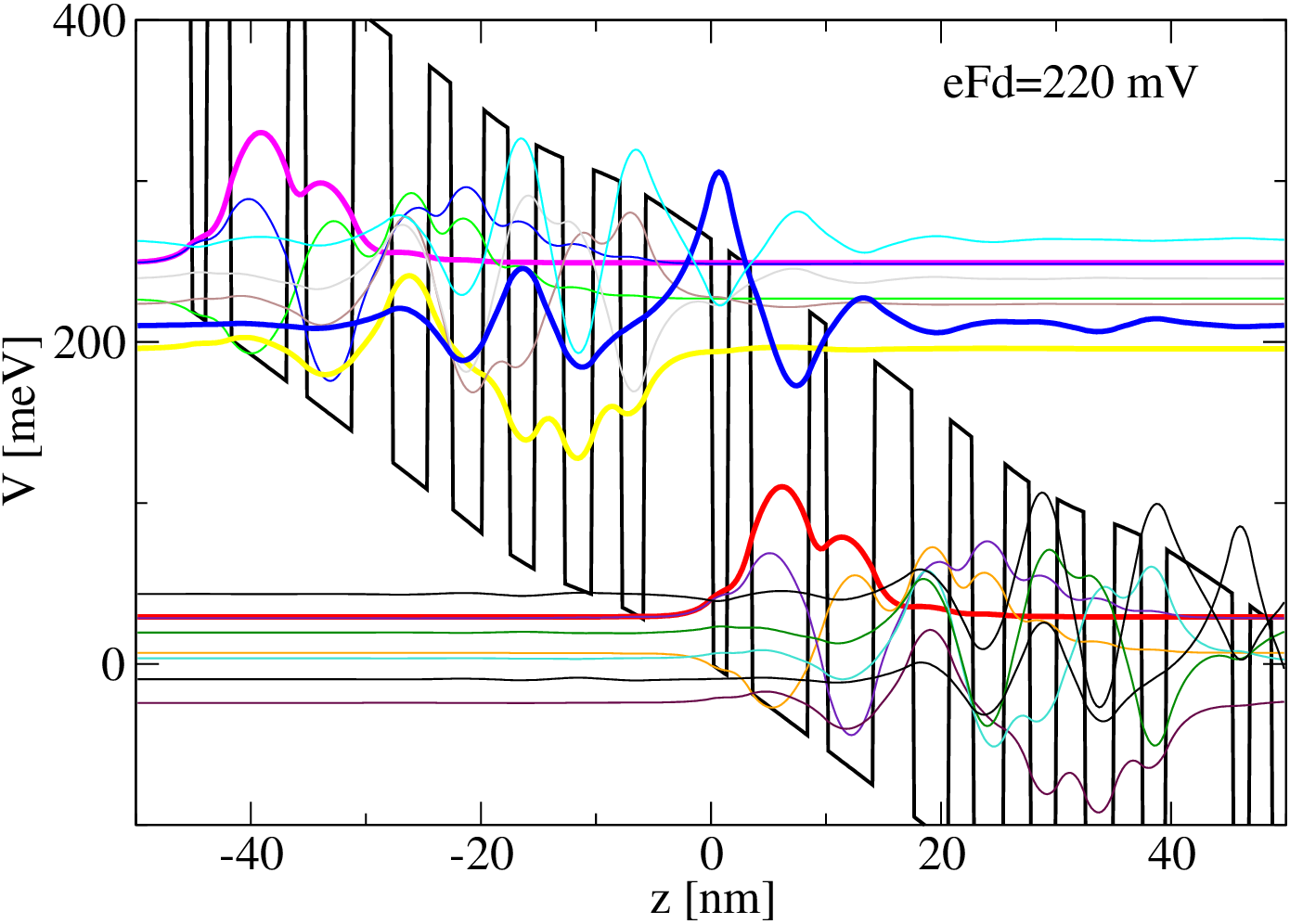}
\caption[]{Conduction band offset including mean field potential
and Wannier functions
for two different electric fields for the sample of  \cite{SIR98}} 
\label{FigWannierFunc}
\end{figure}

We perform our calculation using a basis of Wannier functions. These functions
are shown in Fig.~\ref{FigWannierFunc} for zero bias and
an operating field of 220 mV per period for the sample used in
 \cite{SIR98}. 
While the spatial structure
of the Wannier functions
does not change with bias, their energetic position 
is affected both by the external field and the mean field which
is evaluated self consistently. From Fig.~\ref{FigWannierFunc}
we see, that the mean field almost vanishes at operating conditions
as the electrons are mainly in the injector region where
the doping is also located.
The energy levels of the Wannier functions bunch at the operating field
indicating the strong coupling between the functions enabling transport
through the structure.

\begin{figure}[t]
\includegraphics[width=\textwidth]{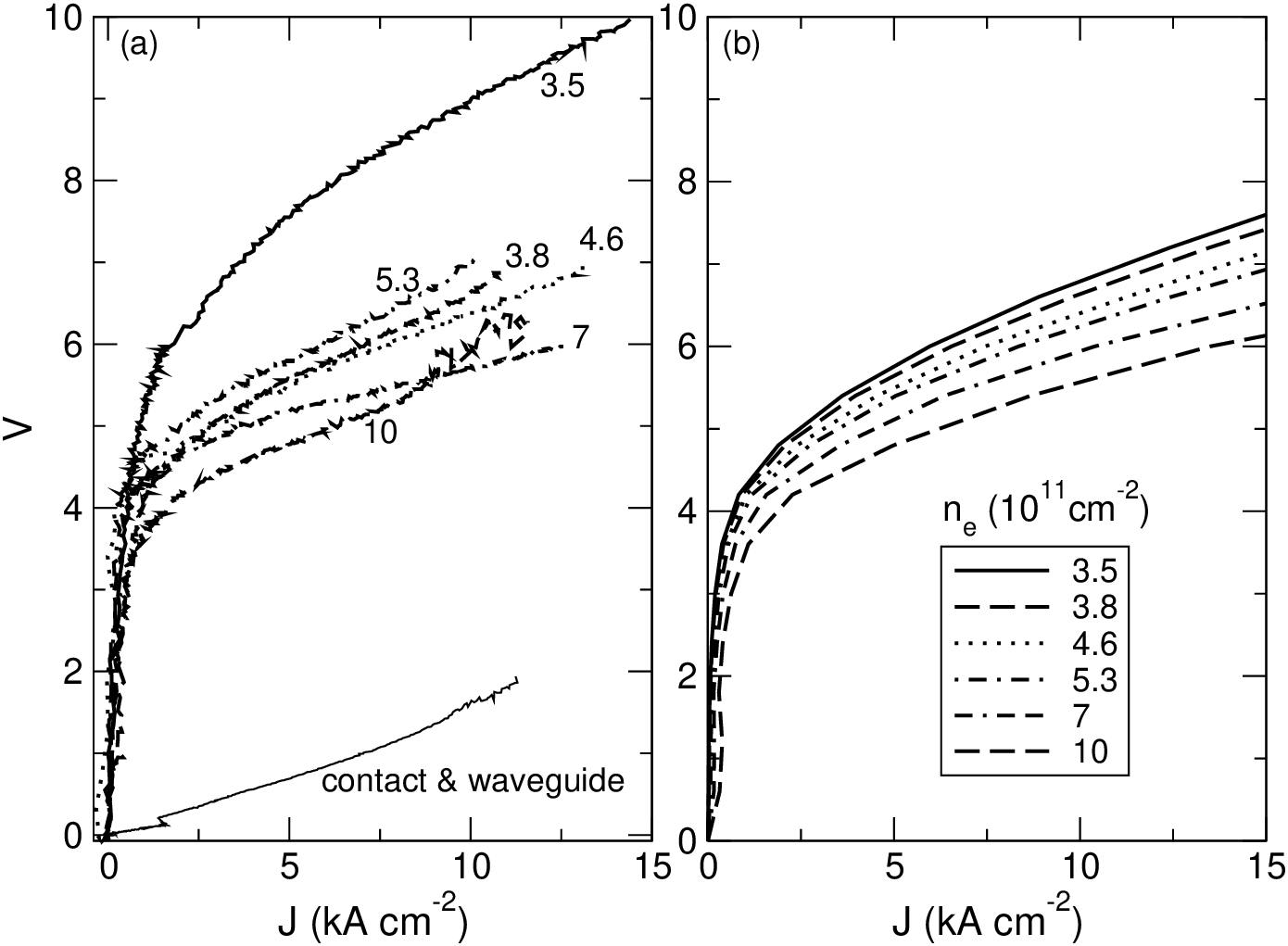}
\caption[]{Current-voltage characteristics for 
different doping densities for the structures of  \cite{GIE03}
at $T=77$ K.
{\bf (a)} Experimental data from M. Giehler (PDI Berlin) where
the thin line refers to a structure grown without the QCL structure,
thus providing an estimate for contact effects.
{\bf (b)} Theoretical result for $\gamma_{\rm imp}=5$ meV
using the bias $U=NFd$, where the QCL 
structure consists of $N=30$ periods }
\label{FigKennDope}
\end{figure}

In Fig.~\ref{FigKennDope} the current-voltage characteristic
is shown for different doping densities $N_D$ per period. 
The theoretical result exhibits 
a monotonic increase of the current density with doping, showing
that the mean field has no dramatic influence on the transport
behavior in these structures. We find good agreement with the
experimental data except for $N_D=3.5\times 10^{11}/\mathrm{cm}^2$ and 
$N_D=5.3\times 10^{11}/\mathrm{cm}^2$ where the experiment
exhibits a significantly higher bias drop. The difference
is of the same order as the bias of a test structure containing
only contact layers and the waveguides, albeit it is not clear
why this additional bias drop is only present in some samples.

\section{Comparing $J_0$ and $J_{\textrm{scatt}}$}

\begin{figure}[t]
\includegraphics[width=0.465\textwidth]{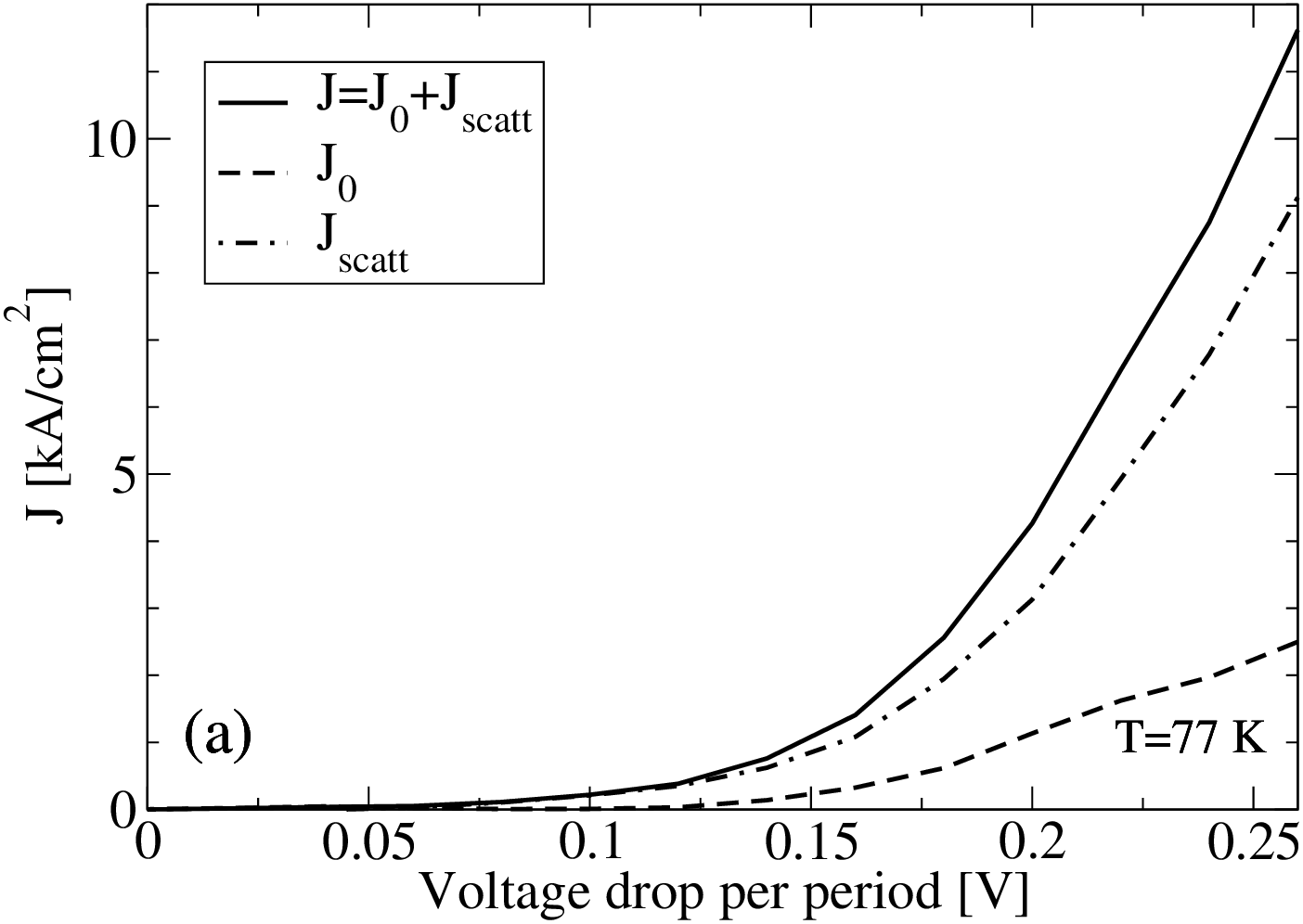}
\hfill
\includegraphics[width=0.50\textwidth]{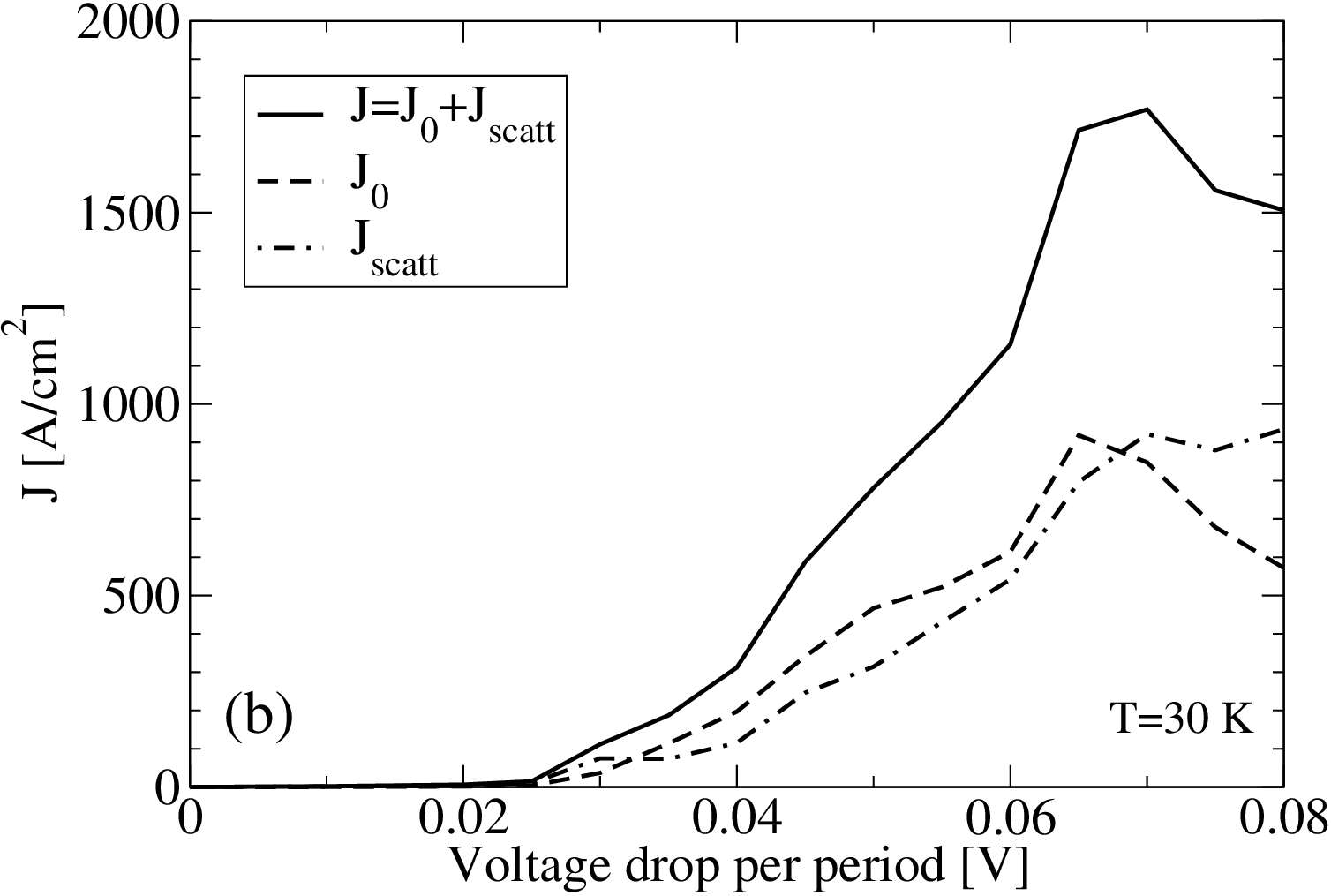}
\caption[]{Contributions to the current density
for the samples of  \cite{SIR98} (a) and  \cite{WIL03} (b).
$\gamma_{\rm imp}=0$ was used in the calculations}
\label{FigJcompare}
\end{figure}

In Fig.~\ref{FigJcompare}(a,b) we show the different contributions to the 
current evaluated for the structures of  \cite{SIR98} and 
 \cite{WIL03}, respectively. Both current-field
relations are in reasonable agreement with the respective
experimental results. (The data of  \cite{WIL03} only
extends to $Fd \approx 70$ mV,
therefore there is no verification of the current peak.)
While $J_{\textrm{scatt}}$ dominates the behavior in
Fig.~\ref{FigJcompare}a, both the contributions
of  $J_{0}$ and $J_{\textrm{scatt}}$ are important in 
Fig.~\ref{FigJcompare}b.

In the following we want to discuss the role of the 
two current contributions with respect to the 
use of semiclassical approaches:
As the expressions for $J_0$ and $J_{\textrm{scatt}}$ are invariant to
unitary transformations of the basis states,
they can be evaluated in arbitrary basis sets.
A special basis set is given by the energy
eigenstates $\varphi_{\mu}(z)$ obtained by diagonalizing $\hat{H}^o$
(including the mean field),
which will be used in the following argumentation.

The semiclassical theories used in \cite{DON01,IOT01a} 
imply that the density matrix is diagonal in the energy
eigenstates,
i.e., $\rho_{\mu{\bf k}\mu'{\bf k}}=\delta_{\mu\mu'}f_{\mu}({\bf k})$.
In this basis the diagonal elements of $W_{\mu\mu'}$ in Eq.~(\ref{EqW}) vanish
and thus $J_0$ becomes zero in the semiclassical approximation.

In the semiclassical approximation 
the Green functions in the basis of energy eigenstates
are given by
\begin{eqnarray}
G^{\rm ret/adv}_{\mu,\nu}({\bf k},E)
&\approx& \mp \pi\imai  \delta_{\mu,\nu}\delta(E-E_{\mu})\\
G^{<}_{\mu,\nu}({\bf k},E)
&\approx& 2\pi\imai \delta_{\mu,\nu}\delta(E-E_{\mu})f_{\mu}({\bf k})
\end{eqnarray}
Then we find from Eq.~(\ref{EqJscattGF})
\begin{equation}\begin{split}
J_{\rm scatt}
 \approx \frac{2 e}{V\hbar} 
\sum_{\mu{\bf k}}\Big\{&
\frac{\imai}{2}\Sigma^{<\, z}_{\mu\mu} ({\bf k},E_{\mu})
+\imai f_{\mu}({\bf k})\Sigma^{{\rm ret}\, z}_{\mu\mu} ({\bf k},E_{\mu})\\
&  -\sum_{\nu}
z_{\mu\nu}\left[\frac{\imai}{2}\Sigma^{<}_{\nu\mu} ({\bf k},E_{\mu})
+\imai f_{\mu}({\bf k}) \Sigma^{\rm ret}_{\nu\mu} ({\bf k},E_{\mu})
\right]
\Big\}
\end{split}\end{equation}
In semiclassical approximation the 
self-energies are related to the scattering probabilities 
$R_{\mu'{\bf k}'\to \mu{\bf k}}$ as follows
\[
\Sigma^{<}_{\mu\mu} ({\bf k},E_{\mu})=
\imai \hbar \sum_{\mu'{\bf k}'} 
f_{\mu'}({\bf k}')R_{\mu'{\bf k}'\to \mu{\bf k}}\, , \quad
\Sigma^{\textrm{ret}}_{\mu\mu} ({\bf k},E_{\mu})=
-\frac{\imai}{2} \hbar \sum_{\mu'{\bf k}'} 
R_{\mu{\bf k}\to \mu'{\bf k}'}
\]
and the quantities $\Sigma^{<\, z}$ and $\Sigma^{\textrm{ret}\, z}$ 
contain an additional factor $z_{\mu'\mu'}$.
This provides us with
\begin{equation}
J_{\rm scatt}
 \approx \frac{2 e}{V} 
\sum_{\mu{\bf k}\mu'{\bf k}'}
R_{\mu{\bf k}\to \mu'{\bf k}'}(z_{\mu'\mu'}-z_{\mu\mu})
\label{EqSemiclassical}
\end{equation}
which is the semiclassical expression for the current 
density\footnote{Terms of the form 
$z_{\mu\nu}\hat{O}_{\nu{\bf k},\mu'{\bf k}'}G_{\mu'\mu'}({\bf k})
\hat{O}_{\mu'{\bf k}',\mu{\bf k}} $
have been neglected for $\nu\neq \mu$ here. Their implication is
not clear yet.}.
Therefore the entire current is contained in $J_{\rm scatt}$ 
in the semiclassical approximation. 

For the special case of the structure considered in \cite{SIR98},
the density matrix is approximately diagonal in the basis of
energy eigenstates\footnote{Note that $J_0$ also vanishes if
the density matrix is diagonal in a basis of real states,
which are not energy eigenstates. But then $J_{\rm scatt}$ no longer
corresponds to the semiclassical result.}
implying that $J_0\to 0$ and $J_{\rm scatt}$ 
is well approximated by the semiclassical expression
(\ref{EqSemiclassical}). This expectation is supported by 
Fig.~\ref{FigJcompare}a and the findings of \cite{IOT01a}.
In contrast, $J_0$ is an important
contribution for  the THz-laser of \cite{WIL03}, which
contains only 4 barriers per period, see Fig.~\ref{FigJcompare}b. 
Therefore it is questionable if semiclassical approaches work here.

\section{Gain and Absorption Spectra}
\begin{figure}[t]\sidecaption
\includegraphics[width=0.6\textwidth]{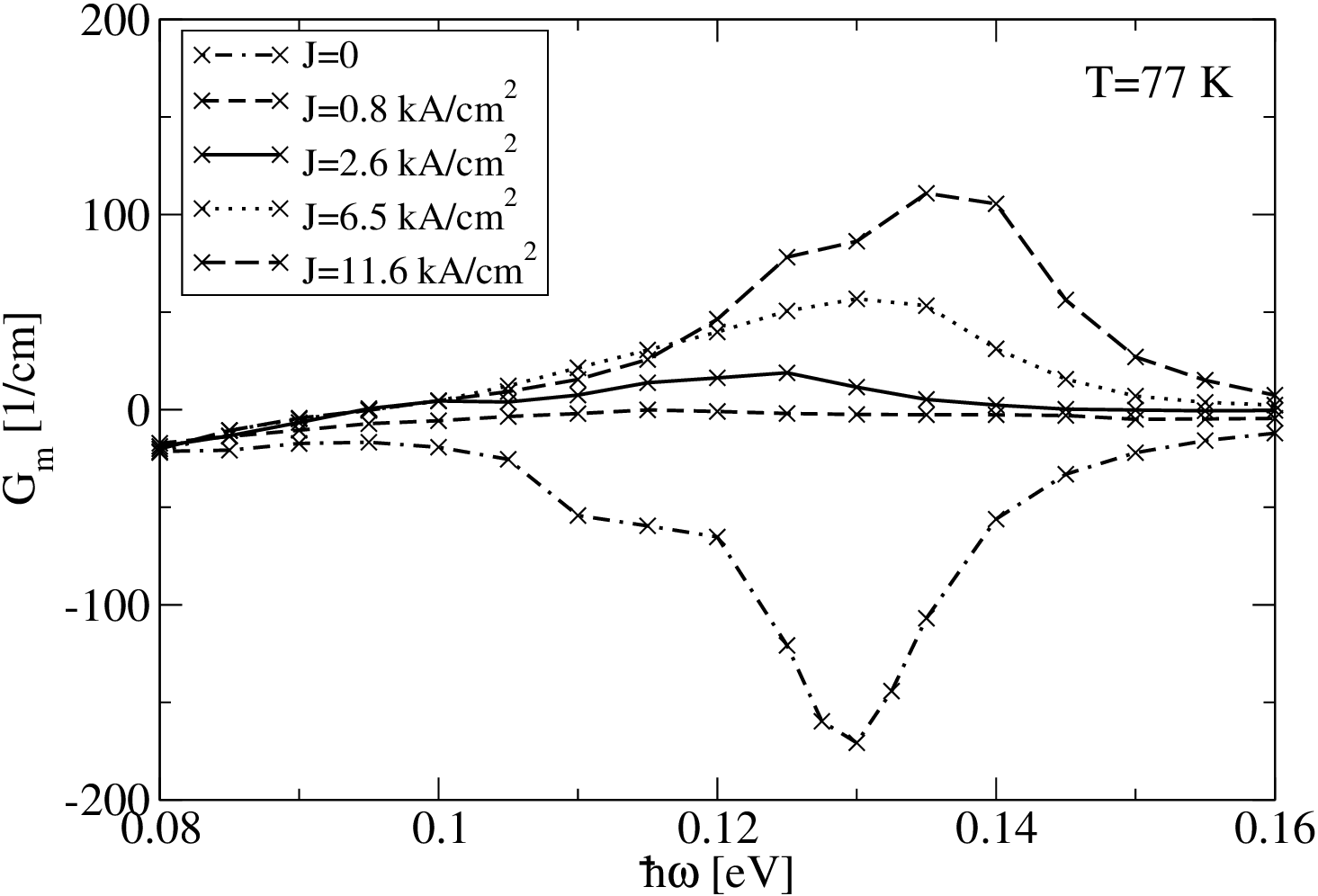}
\caption[]{Gain spectrum for the sample of  \cite{SIR98}
using $\gamma_{\rm imp}=0$. (From  \cite{WAC02b})}
\label{FigGainSirt}
\end{figure}

The general evaluation of gain spectra within the quantum transport
model used here was described in detail
in  \cite{WAC02b}. The key idea is to evaluate the complex 
susceptibility $\chi(\omega)$ which is related to the optical
absorption coefficient at a frequency $\omega$ 
via \cite{JAC98a}
\begin{equation}
\alpha(\omega)= \frac{\omega}{c} \frac{\Im\{\chi(\omega)\}}{n_B},
\label{EqgainbyXi}
\end{equation}
where $n_B$ is the background refractive index and $c$
is the speed of light. Figure~\ref{FigGainSirt} shows
the gain spectrum for the sample of  \cite{SIR98}.
At zero current we find strong absorption due to transitions
in the active region, which vanishes already for small currents
as the carriers are transfered to the injector region. 
Pronounced gain around $\hbar\omega=130$ meV sets in for current
densities of several kA/cm$^2$. The height and width of the
gain spectrum is in good agreement with the findings of 
 \cite{EIC00}.

Within the semiclassical approximation the susceptibility is given
by \cite{Hau93}
\begin{equation}
{\Im}\{\chi(\omega)\} = \sum_{\mu\nu{\bf k}}
\frac{|\wp_{\mu\nu}|^2}{\epsilon_0V}
\frac{[f_{\mu}({\bf k}) - f_{\nu}({\bf k})](\Gamma_{\nu}+\Gamma_{\mu})}
{(E_{\nu}-E_{\mu}-\hbar\omega)^2+(\Gamma_{\nu}+\Gamma_{\mu})^2/4}\, ,
\label{EqXisimp}
\end{equation}
where $\Gamma_{\nu}$ is the FWHM of 
${\Im}\{G^{\rm ret}_{\nu\nu}({\bf k},E)\}$ and 
$\wp_{\mu\nu}=ez_{\mu\nu}$ is the dipole matrix element.
In  \cite{LEE02a} it was shown that this  semiclassical 
approach gives reasonable results compared with the quantum model
for the structure of  \cite{SIR98}.

In these approaches the influence
of electron-electron interaction was totally neglected. 
Here we study the influence of many-particle corrections within the 
Hartree Fock approximation on the gain
spectrum. 
The susceptibility is decomposed by
 \[
\chi(\omega) = 2\mbox{(for spin)}
\sum_{\mu,\nu,{\bf k}} \wp_{\mu \nu}\chi_{\nu,\mu}({\bf k},\omega)
\]
where the susceptibility functions $\chi_{\nu,\mu}({\bf k},\omega)$
between the eigenstates $\nu$ and $\mu$
are determined by the equation 
\begin{equation}\begin{split}
\wp_{\nu \mu} ({\bf k}) & \left( f_{\nu}({\bf k})-f_{\mu}({\bf k}) \right)=\\
& \hbar \left(  \omega - e_{\nu}({\bf k}) + e_{\mu}({\bf k}) +\imai (\Gamma_{\mu}+\Gamma_{\nu})/2 \right) 
\chi_{\nu\mu}({\bf k},\omega)& \\ 
&+\left( f_{\nu}({\bf k})-f_{\mu}({\bf k}) \right) \; 2 \;
V \left( \begin{array}{c} \nu \; \mu \; \mu \; \nu \\ 0 \end{array} \right) \; \sum_{{\bf k}'} \chi_{\nu\mu}({\bf k}',\omega)\\
&-\left( f_{\nu}({\bf k})-f_{\mu}({\bf k}) \right) \;
\sum_{{\bf k}'} \chi_{\nu\mu}({\bf k}',t) \;  
V \left( \begin{array}{c} \nu \; \nu \; \mu \; \mu \\ {\bf k} -{\bf k}' 
\end{array} \right)\, .
\label{EqHarFock}
\end{split}\end{equation}
Equation (\ref{EqHarFock}) reduces, in the equilibrium case, with only two
isolated subbands of idealized quantum well subbands where 
phenomenological dephasing characterizes the broadening, to Eq.~(5) of 
 \cite{CHU92}.
The bare Coulomb interaction and renormalized energies which appear above
are given by
\[
V \left(\begin{array}{c} \mu \; \nu \; \alpha \; \beta \\
{\bf k}-{\bf k}' \end{array} \right) = 
\int \d z \, \d z'\,
\phi^{\ast}_{\mu} (z) 
\phi_{\nu} (z) 
\frac{e^2\, \e^{-|{\bf k}-{\bf k}' | | z - z' | }}
{2\epsilon_r\epsilon_0A|{\bf k}-{\bf k}' |} 
\phi^{\ast}_{\alpha} (z')
\phi_{\beta} (z')
\]
with the normalization area $A$ and 
\[
\hbar e_{\nu}({\bf k}) = E_{\nu}({\bf k}) 
- \sum_{{\bf k}'} f_{\nu}({\bf k}') 
\; V \left(\begin{array}{c}  \nu \; \nu \; \nu \; \nu \\ 
{\bf k}-{\bf k}' \end{array} \right)  
+ \sum_{{\bf k}'} f_{\nu}({\bf k}') 
\; V \left( \begin{array}{c} \nu \; \mu \; \mu \; \nu \\ 
{\bf k}-{\bf k}' \end{array}  \right)\, .
\]

\begin{figure}[t]\sidecaption
\includegraphics[width=0.6\textwidth]{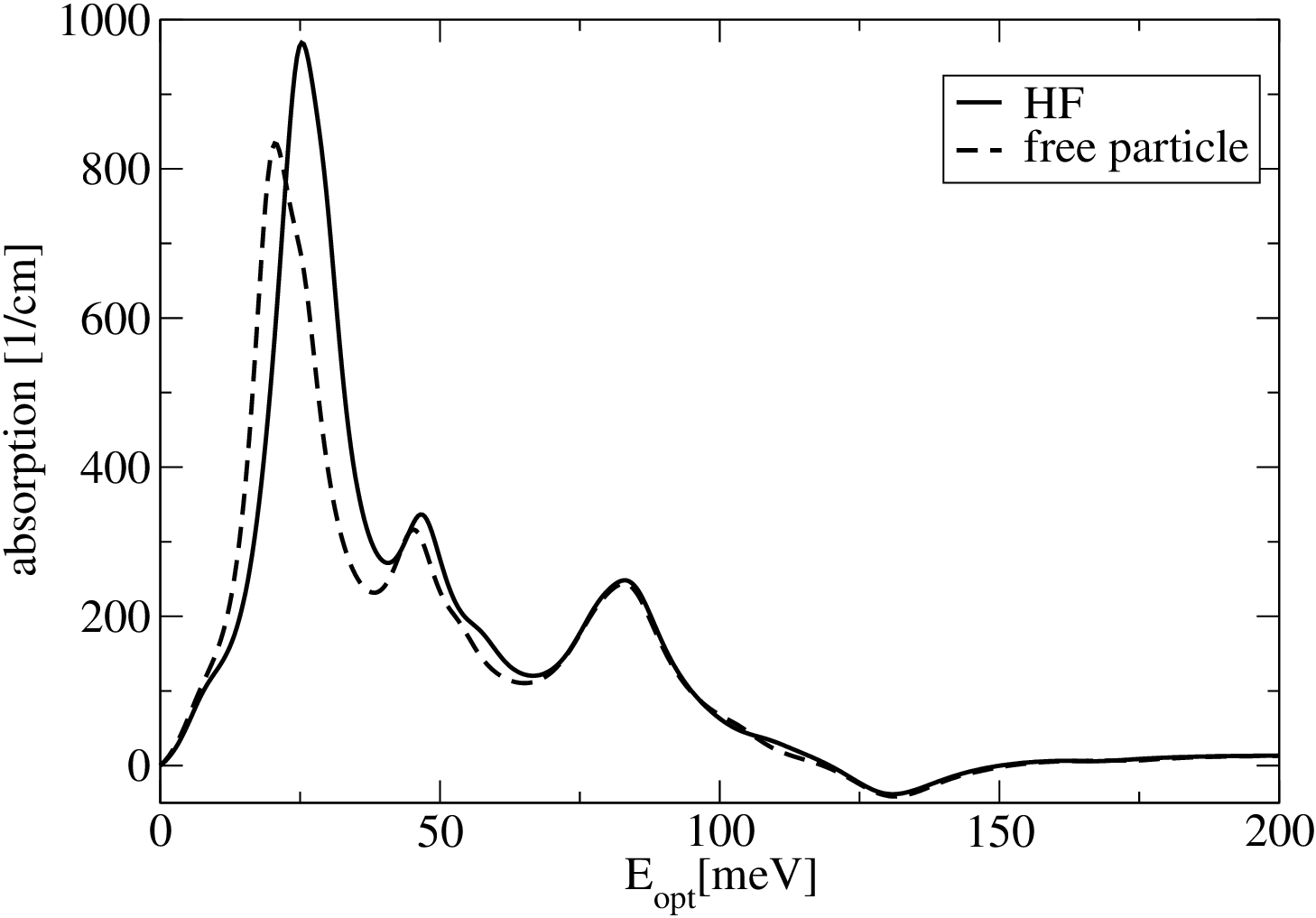}
\caption[]{Absorption spectrum for the sample from Fig.~\ref{FigKennDope}
with $N_d=3.8\times 10^{11}/\mathrm{cm}^2$ at an operating field
of $Fd=0.2$ V. The dashed line gives the
result from Eq.~(\ref{EqXisimp}). The full line includes
Coulomb corrections according to Eq.~(\ref{EqHarFock})}
\label{FigAbsHF}
\end{figure}

The second term on the right-hand side of (\ref{EqHarFock}) 
gives rise to the depolarization
shift \cite{AND82,HEL99a}, while the last term (exchange contribution) 
is analogous to the 
excitonic coupling term in interband transitions \cite{PER94}.
Figure~\ref{FigAbsHF} shows the absorption 
spectra. The inclusion of many-particle corrections 
yields a blue shift of about 5 meV for the low frequency absorption
peak and a slight red shift for the gain peak around 130 meV.

\section{Discussion}
The impact of quantum effects on transport and gain in quantum cascade
lasers have been examined. In the evaluation of
the current two different terms, $J_0\propto [\hat{H}_o,\hat{z}]$ and 
$J_{\rm scatt}\propto [\hat{H}_{\rm scatt},\hat{z}]$ appear.
In the semiclassical approximation, where the density matrix is assumed
to be diagonal in the basis of energy eigenstates, 
$J_{\rm scatt}$ carries the entire current.
Our quantum transport calculations show that $J_{\rm scatt}$ dominates
the behavior for the prototype sample of  \cite{SIR98},
which has been frequently studied, thus justifying the
semiclassical approaches in \cite{DON01,IOT01a}.
On the other hand the current $J_0$, resulting from nondiagonal
elements in the density matrix, shows strong contributions
for the THz-laser of  \cite{WIL03}.
Nevertheless, it is not clear by now in how far the assumption
of diagonal self-energies in the Wannier basis affects this
behavior. Ongoing work is focused towards the inclusion
of the full matrix structure in the self-energies.

The many particle effects are extremely dependent on the structure,
since its design determines the 
actual electronic overlap and subband occupation.
For the case considered 
here, the gain spectra are hardly modified by the electron-electron 
interactions within the Hartree-Fock approximations, while
a significant depolarization shift occurs for the low
frequency absorption.

Helpful discussions with M.~Giehler, H.T.~Grahn, A.~Knorr, L.~Schrottke, 
and M.~W{\"o}rner as well as financial support by DFG within  FOR394
is gratefully acknowledged.

\appendix
\section*{Appendix}

Similarly to \cite{LEE02a} the scattering current (\ref{EqJscatt})
can be evaluated in the following way:
We define the contour-ordered
Green function (superscript $c$)
\[\begin{split}
F^c_{\alpha{\bf k}}(\tau_1,\tau_2) 
=\sum_{\gamma\beta {\bf k'}}
 -i\hat{T}_c\{&
\langle \hat{O}_{\alpha{\bf k},\beta{\bf k'}}(\tau_1)
z_{\beta\gamma} \hat{a}_{\gamma{\bf k'}}(\tau_1) 
\hat{a}^\dagger_{\alpha {\bf k}}(\tau_2) \\
&- 
z_{\alpha\beta}\hat{O}_{\beta{\bf k},\gamma{\bf k'}}(\tau_1)
\hat{a}_{\gamma{\bf k'}}(\tau_1) 
\hat{a}^\dagger_{\alpha {\bf k}}(\tau_2)\rangle\}\, .
\end{split}\]
They are evaluated in the Dirac representation (with index $D$) 
\[\begin{split}
F^c_{\alpha{\bf k}}(\tau_1,\tau_2) 
=\sum_{\gamma\beta {\bf k'}}
 -i\hat{T}_c\Big\langle&
\e^{\int\d \tau \frac{1}{\imai\hbar}\hat{H}_{\rm scatt}(\tau)}\Big\{
\hat{O}^D_{\alpha{\bf k},\beta{\bf k'}}(\tau_1)
Z_{\beta\gamma} \hat{a}^D_{\gamma{\bf k'}}(\tau_1) 
\hat{a}^{D\dagger}_{\alpha {\bf k}}(\tau_2) \\
&- Z_{\alpha\beta}\hat{O}^D_{\beta{\bf k},\gamma{\bf k'}}(\tau_1)
\hat{a}^D_{\gamma{\bf k'}}(\tau_1) 
\hat{a}^{D\dagger}_{\alpha {\bf k}}(\tau_2)\Big\}
\Big\rangle
\end{split}\]
The lowest order non-vanishing terms of the expansion gives
\[\begin{split}
F^c_{\alpha{\bf k}}&(\tau_1,\tau_2) 
\approx\sum_{\gamma\beta {\bf k'}}
\frac{1}{\hbar} \sum_{\delta\epsilon} \int d\tau\,\\
\Big\langle&
\hat{O}_{\alpha{\bf k},\beta{\bf k'}}(\tau_1)z_{\beta\gamma}
G^{c0}_{\gamma,\delta}({\bf k'};\tau_1,\tau)
\hat{O}_{\delta{\bf k'},\epsilon{\bf k}}(\tau)
G^{c0}_{\epsilon,\alpha}({\bf k};\tau,\tau_2) \\ 
&  -z_{\alpha\beta}
\hat{O}_{\beta{\bf k},\gamma{\bf k'}}(\tau_1)
G^{c0}_{\gamma,\delta}({\bf k'};\tau_1,\tau)
\hat{O}_{\delta{\bf k'},\epsilon{\bf k}}(\tau)
G^{c0}_{\epsilon,\alpha}({\bf k};\tau,\tau_2)\Big\rangle
\end{split} \]
with the bare Green functions
$G^{c0}_{\alpha,\gamma}({\bf k};\tau_1,\tau)
=-i\hat{T}_c\{ \langle \hat{a}^D_{\alpha{\bf k}}(\tau_1) 
\hat{a}^{D\dagger}_{\gamma{\bf k}}(\tau)\rangle\}$.
In order to be consistent with the perturbation expansion in
the Green functions, further terms are taken into account, which
replace the bare Green functions by the full Green functions. Then we 
find
\[\begin{split}
F^c_{\alpha{\bf k}}(\tau_1,\tau_2) 
\approx
\frac{1}{\hbar} \sum_{\epsilon} \int d\tau\,
\Big[& \Sigma^{c\, z}_{\alpha\epsilon} ({\bf k};\tau_1,\tau)
G^{c}_{\epsilon,\alpha}({\bf k};\tau,\tau_2) \\ 
&  -\sum_{\beta}
z_{\alpha\beta}\Sigma^{c}_{\beta\epsilon}({\bf k};\tau_1,\tau)
G^{c}_{\epsilon,\alpha}({\bf k};\tau,\tau_2)\Big]
\end{split} \]
with
\begin{equation}
\Sigma^{c\, z}_{\alpha\epsilon} ({\bf k};\tau_1,\tau)
=\sum_{\gamma\beta\delta {\bf k'}}\langle
\hat{O}_{\alpha{\bf k},\beta{\bf k'}}(\tau_1)z_{\beta\gamma}
G^{c}_{\gamma,\delta}({\bf k'};\tau_1,\tau)
\hat{O}_{\delta{\bf k'},\epsilon{\bf k}}(\tau)\rangle
\end{equation}
where the averaging refers to the phonon bath for 
phonon scattering. Thus, in Born approximation 
the self energies $\Sigma^{z}$ are 
given the the usual functionals for the self-energies $\pmb{\Sigma}({\bf G})$ 
where the Green functions ${\bf G}$ are replaced by ${\bf Z}\cdot{\bf G}$
in matrix notation.
Using Langreth rules and changing to the energy 
representation $F^<_{\alpha{\bf k}}(t,t)$ can be
inserted in Eq.~(\ref{EqJscatt}) yielding the
final expression
\begin{equation}\begin{split}
J_{\rm scatt}=&\frac{2e}{V\hbar} 
\sum_{\alpha{\bf k}}F^<_{\alpha{\bf k}}(t,t)\\ 
=& \frac{2e}{V\hbar} 
\sum_{\alpha{\bf k}}
\int\frac{\d E}{2\pi}\Big\{\\
&\sum_{\epsilon}\left[\Sigma^{<\, z}_{\alpha\epsilon} ({\bf k},E)
G^{\rm adv}_{\epsilon,\alpha}({\bf k},E)
+\Sigma^{{\rm ret}\, z}_{\alpha\epsilon} ({\bf k},E)
G^{<}_{\epsilon,\alpha}({\bf k},E)\right]\\ 
&  -\sum_{\epsilon\beta}
z_{\alpha\beta}\left[\Sigma^{<}_{\beta\epsilon} ({\bf k},E)
G^{\rm adv}_{\epsilon,\alpha}({\bf k},E)
+\Sigma^{\rm ret}_{\beta\epsilon} ({\bf k},E)
G^{<}_{\epsilon,\alpha}({\bf k},E)\right]
\Big\}
\label{EqJscattGF}
\end{split}\end{equation}
to evaluate $J_{\rm scatt}$.


\begin{thebibliography}{10}

\bibitem{FAI94a}
J. Faist, F. Capasso, D.~L. Sivco, C. Sirtori, A.~L. Hutchinson, and A.~Y. Cho,
  Science {\bf 264},  553  (1994).

\bibitem{BEC02a}
M. Beck, D. Hofstetter, T. Aellen, J. Faist, U. Oesterle, M. Ilegems, E. Gini,
  and H. Melchior, Science {\bf 295},  301  (2002).

\bibitem{KOE02}
R. K{\"o}hler, A. Tredicucci, F. Beltram, H.~E. Beere, E.~H. Linfield, A.~G.
  Davies, D.~A. Ritchie, R.~C. Iotti, and F. Rossi, Nature {\bf 417},  156
  (2002).

\bibitem{SIR98}
C. Sirtori, P. Kruck, S. Barbieri, P. Collot, J. Nagle, M. Beck, J. Faist, and
  U. Oesterle, Appl.~Phys.~Lett. {\bf 73},  3486  (1998).

\bibitem{SCA97}
G. Scamarcio, F. Capasso, J. Faist, C. Sirtori, D.~L. Sivco, A.~L. Hutchinson,
  and A.-Y. Cho, Appl.~Phys.~Lett. {\bf 70},  1796  (1997).

\bibitem{STR99}
G. Strasser, S. Gianordoli, L. Hvozdara, W. Schrenk, K. Unterrainer, and E.
  Gornik, Appl.~Phys.~Lett. {\bf 75},  1345  (1999).

\bibitem{WAN01}
M.~C. Wanke, F. Capasso, C. Gmachl, A. Tredicucci, D.~L. Sivco, A.~L.
  Hutchinson, S.-N.~G. Chu, and A.~Y. Cho, Appl.~Phys.~Lett. {\bf 78},  3950
  (2001).

\bibitem{ULB02}
N. Ulbrich, G. Scarpa, G. B{\"o}hm, G. Abstreiter, and M. Amann,
  Appl.~Phys.~Lett. {\bf 80},  4312  (2002).

\bibitem{WIL03}
B.~S. Williams, H. Callebaut, S. Kumar, Q. Hu, and J.~L. Reno,
  Appl.~Phys.~Lett. {\bf 82},  1015  (2003).

\bibitem{CAP96}
F. Capasso, J. Faist, and C. Sirtori, J.~Math.~Phys. {\bf 37},  4775  (1996).

\bibitem{PAU98}
D. Paulavi\v{c}ius, V. Mitin, and M.~A. Stroscio, J.~Appl.~Phys. {\bf 84},
  3459  (1998).

\bibitem{SLI99}
S. Slivken, V.~I. Litvinov, M. Razeghi, and J.~R. Meyer, J.~Appl.~Phys. {\bf
  85},  665  (1999).

\bibitem{HYL96}
P. Hyldgaard and J.~W. Wilkins, Phys.~Rev.~B {\bf 53},  6889  (1996).

\bibitem{HAR99}
P. Harrison, Appl.~Phys.~Lett. {\bf 75},  2800  (1999).

\bibitem{APA01}
V.~M. Apalkov and T. Chakraborty, Appl.~Phys.~Lett. {\bf 78},  1973  (2001).

\bibitem{TOR99}
S. Tortora, F. Compagnone, A. {Di Carlo}, P. Lugli, M.~T. Pellegrini, M.
  Troccoli, and G. Scamarcio, Physica~B {\bf 272},  219  (1999).

\bibitem{IOT00}
R.~C. Iotti and F. Rossi, Appl.~Phys.~Lett. {\bf 76},  2265  (2000).

\bibitem{IOT01}
R.~C. Iotti and F. Rossi, Appl.~Phys.~Lett. {\bf 78},  2902  (2001).

\bibitem{DON01}
K. Donovan, P. Harrison, and R.~W. Kelsall, J.~Appl.~Phys. {\bf 89},  3084
  (2001).

\bibitem{IOT01a}
R.~C. Iotti and F. Rossi, Phys.~Rev.~Lett. {\bf 87},  146603  (2001).

\bibitem{WAC01a}
A. Wacker,  in {\em Advances in Solid State Phyics}, edited by B. Kramer
  (Springer, Berlin, 2001), p.\ 199.

\bibitem{LEE02a}
S.-C.~Lee and A. Wacker, Phys.~Rev.~B {\bf 66},  245314  (2002).

\bibitem{KUH98}
T. Kuhn,  in {\em Theory of Transport Properties of Semiconductor
  Nanostructures}, edited by E. Sch{\"o}ll (Chapman and Hall, London, 1998).

\bibitem{KAD62}
L.~P. Kadanoff and G. Baym, {\em Quantum Statistical Mechanics} (Benjamin, New
  York, 1962).

\bibitem{KEL65}
L.~V. Keldysh, Sov.~Phys.~JETP {\bf 20},  1018  (1965), [Zh.~Eksp.~Theor.~Fiz.\
  {\bf 47}, 1515 (1964)].

\bibitem{HAU96}
H. Haug and A.-P. Jauho, {\em Quantum Kinetics in Transport and Optics of
  Semiconductors} (Springer, Berlin, 1996).

\bibitem{WAC02}
A. Wacker, Phys.~Rep. {\bf 357},  1  (2002).

\bibitem{GIE03}
M. Giehler, R. Hey, H. Kostial, S. Cronenberg, T. Ohtsuka, L. Schrottke, and
  H.~T. Grahn, Appl.~Phys.~Lett. {\bf 82},  671  (2003).

\bibitem{WAC02b}
A. Wacker, Phys.~Rev.~B {\bf 66},  085326  (2002).

\bibitem{JAC98a}
J.~D. Jackson, {\em Classical Electrodynamics}, 3 ed. (John Wiley \& Sons, New
  York, 1998).

\bibitem{EIC00}
F. Eickemeyer, R.~A. Kaindl, M. Woerner, T. Elsaesser, S. Barbieri, P. Kruck,
  C. Sirtori, and J. Nagle, Appl.~Phys.~Lett. {\bf 76},  3254  (2000).

\bibitem{Hau93}
H. Haug and S.~W. Koch, {\em Quantum Theory of the Optical and Electronic
  Properties of Semiconductors}, 2 ed. (World Scientific, Singapore, 1993).

\bibitem{CHU92}
S.~L. Chuang, M.~S.~C. Luo, S. Schmitt-Rink, and A. Pinczuk, Phys.~Rev.~B {\bf
  46},  1897  (1992).

\bibitem{AND82}
T. Ando, A.~B. Fowler, and F. Stern, Rev.~Mod.~Phys. {\bf 54},  437  (1982).

\bibitem{HEL99a}
M. Helm,  in {\em Intersubband Transitions in Quantum Wells: Physics and Device
  Applications}, edited by E.~R. Weber and R.~K. Willardson 
  (Academic Press, 1999), Vol.~62, p.\ 1.

\bibitem{PER94}
M.~F. {Pereira, Jr.}, R. Binder, and S.~W. Koch, Appl.~Phys.~Lett. {\bf 64},
  279  (1994).

\end{thebibliography}

\end{document}